\documentclass[nobib]{tufte-handout}

\usepackage[utf8]{inputenc}
\usepackage[T1]{fontenc}
\usepackage{amsmath, amsfonts, amssymb}
\usepackage{graphicx}
\usepackage{xcolor}
\usepackage{algorithm}
\usepackage{algpseudocode}
\usepackage{booktabs}
\usepackage{hyperref}

\usepackage{tabularx}
\hypersetup{
    colorlinks=true,
    linkcolor=blue,
    urlcolor=blue,
}

\algnewcommand{\Continue}{\textbf{continue}}

\usepackage{cite}      
\usepackage{url}      
\usepackage{doi}

\usepackage{listings}
\usepackage{hyperref}

\usepackage{cite}

\usepackage{lmodern}

\newcommand{\cmcolors}{\texttt{cm-colors}}

\newcommand{\oklch}{\textsc{oklch}}
\newcommand{\WCAG}{\textsc{WCAG}}

\newcommand{\fnm}[1]{#1}
\newcommand{\sur}[1]{#1}

\title{Perceptually-Minimal Color Optimization for Web Accessibility: A Multi-Phase Constrained Approach}

\author{%
  \fnm{Lalitha} \sur{A R} \newline
  \small Email: 24f2006078@ds.study.iitm.ac.in \newline
  \small GitHub: @lalithaar \newline
  \small ORCID: 0009-0001-7466-3531 \newline
}

\date{\today}

\begin{document}

\maketitle

\begin{abstract}
Web accessibility guidelines require sufficient color contrast between text and backgrounds; yet, manually adjusting colors often necessitates significant visual deviation, compromising vital brand aesthetics. We present a novel, multi-phase optimization approach for automatically generating WCAG-compliant colors while \textbf{minimizing perceptual change} to original design choices.

Our method treats this as a \textbf{constrained, non-linear optimization problem}, utilizing the modern \textbf{perceptually uniform OKLCH color space}. Crucially, the optimization is constrained to preserve the original hue ($\text{h}$) of the color, ensuring that modifications are strictly limited to necessary adjustments in lightness ($\text{L}$) and chroma ($\text{C}$). This is achieved through a three-phase sequence: binary search, gradient descent, and progressive constraint relaxation.

Evaluation on a dataset of 10,000 procedurally generated color pairs demonstrates that the algorithm successfully resolves accessibility violations in $77.22\%$ of cases, with $88.51\%$ of successful corrections exhibiting \textbf{imperceptible color difference} ($\Delta E_{2000} < 2.0$) as defined by standard perceptibility thresholds. The median perceptual change for successful adjustments is only $0.76\ \Delta E_{2000}$, and the algorithm achieves this with a median processing time of $0.876\text{ms}$ per color pair.

The approach demonstrates that accessibility compliance and visual design integrity can be achieved simultaneously through a \textbf{computationally efficient}, perceptually-aware optimization that respects brand identity. The algorithm is publicly implemented in the open-source cm-colors Python library.
\end{abstract}

\section{Introduction}

Web accessibility requirements often create tension between inclusive design and brand aesthetics. The \WCAG\ 2.1\cite{W3C:WCAG21} guidelines require minimum contrast ratios of 4.5:1 for normal text and 3:1 for large text to meet AA compliance standards\footnote{WCAG AAA compliance requires 7:1 for normal text and 4.5:1 for large text, representing the gold standard for accessibility.}. Large-scale studies consistently identify insufficient color contrast as one of the most prevalent accessibility issues, affecting millions of websites \cite{webaim2025, martins2024}. While color contrast checking tools readily identify accessibility violations, existing solutions typically suggest binary fixes that frequently compromise carefully crafted visual identities.

This work provides a technical implementation of the \textbf{Design Harmony} principle from the Comfort Mode Framework \cite{comfort_mode_framework}, which advocates for reconciling accessibility with aesthetic integrity rather than treating them as competing objectives. While the framework establishes the theoretical necessity for this harmony, practical implementation requires sophisticated mathematical approaches.

We present a novel approach that treats accessibility improvement as a constrained optimization problem in perceptually uniform color space. Our method automatically generates \WCAG-compliant colors while minimizing visual disruption to original design intentions. Evaluation on 10,000 color pairs shows the algorithm successfully resolves accessibility violations in 77.22\% of cases with median perceptual change of only 0.76 $\Delta E_{2000}$\footnote{Delta E 2000 is the current CIE standard for perceptual color difference, with values below 2.0 representing barely perceptible changes under normal viewing conditions.}\cite{Luo2001}, while maintaining real-time performance (0.876ms per pair).

The complete implementation, demo, and documentation are available as the open-source \cmcolors\ library\footnote{Available at \cite{A_R_CM-Colors_You_pick} and \url{https://comfort-mode-toolkit.github.io/cm-colors/}}.

\section{Theoretical Foundation}

\subsection{Color Accessibility and Perceptual Uniformity}

Standard RGB color representation creates significant challenges for perceptual optimization. Numerical changes in RGB space do not correspond to equal perceptual changes\cite{fairchild2013color}, making traditional optimization metrics inappropriate for preserving visual aesthetics\footnote{For example, changing RGB(255,87,51) to RGB(250,87,51) appears as a tiny numerical change but may produce significant perceptual differences depending on the specific color region.}. 

We address this fundamental limitation by converting all optimization operations to \oklch\ color space\footnote{OKLCH is based on the 2020 OKLAB color appearance model, designed to be more perceptually uniform than previous attempts like LAB or LUV. The conversion involves: RGB → Linear RGB → XYZ → OKLAB → OKLCH cylindrical coordinates.}:

\begin{itemize}
\item \textbf{L (Lightness)}: Perceptual brightness from 0 (black) to 1 (white)
\item \textbf{C (Chroma)}: Colorfulness/saturation from 0 (gray) to maximum gamut
\item \textbf{H (Hue)}: Color angle in degrees (0-360)
\end{itemize}

\oklch\ space provides perceptually uniform characteristics where equal numerical changes produce equal visual differences, enabling mathematically meaningful optimization of aesthetic preservation while achieving accessibility goals \cite{oklab_model}.

\subsection{Problem Formulation}

We formalize the accessibility-preserving color optimization as a constrained optimization problem\footnote{Breaking this down: (1) Minimize visual difference from original color, (2) Ensure sufficient contrast with background, (3) Keep colors within standard RGB range.}:

\begin{equation}
\begin{aligned}
\min_{c'} \quad & \Delta E_{2000}(c_{original}, c') \\
\text{subject to} \quad & \text{contrast}(c', c_{bg}) \geq \tau \\
& c' \in \text{sRGB gamut}
\end{aligned}
\end{equation}

where $c'$ represents the optimized color, $\tau$ is the target contrast ratio (4.5:1 or 7:1 depending on compliance level), and $\Delta E_{2000}$ quantifies perceptual color difference \cite{cie_de2000}.

This formulation directly implements the Design Harmony principle by minimizing perceptual change (preserving brand aesthetics) while satisfying accessibility constraints (ensuring inclusive access).

\section{Algorithm Design}

Our multi-phase optimization approach\footnote{If you would prefer to read python implementation instead, \href{https://github.com/comfort-mode-toolkit/cm-colors/raw/refs/heads/main/src/cm_colors/core/optimisation.py}{ please check here}} leverages the mathematical structure of the accessibility constraint problem to achieve efficient solutions with minimal brand impact. The algorithm employs three sequential phases with progressive constraint relaxation, designed to prioritize computational speed (Phase 1) and solution quality (Phase 2), governed by a comprehensive control function (Phase 3).

\subsection{Phase 1: Binary Search on Lightness}

Most accessibility violations can be resolved through lightness adjustments alone\footnote{In essence: Systematically adjusts the color's brightness toward the background's opposite—lightening dark text or darkening light text—until sufficient contrast is achieved with minimal visual change.}, as contrast ratio depends primarily on relative luminance differences. We implement precision-matched binary search on the L component of \oklch\ space that not only finds compliant colors but continues optimization to minimize perceptual distance.

\begin{algorithm}[H]
\caption{Binary Search Lightness Optimization ($\text{BinarySearchLightness}$)}
\begin{algorithmic}[1]
\Require $c_{text}$, $c_{bg}$, $\delta_E^{max}$, $\tau_{target}$
\Ensure Optimized color $c'$ or $\emptyset$

\State $(L, C, H) \leftarrow \text{RGB\_to\_OKLCH}(c_{text})$
\State $(L_{bg}, -, -) \leftarrow \text{RGB\_to\_OKLCH}(c_{bg})$
\State $\text{search\_up} \leftarrow L_{bg} < 0.5$ \Comment{Lighten on dark bg, darken on light bg}

\If{search\_up}
    \State $L_{low} \leftarrow L$, $L_{high} \leftarrow 1.0$
\Else
    \State $L_{low} \leftarrow 0.0$, $L_{high} \leftarrow L$
\EndIf

\State $c_{best} \leftarrow \emptyset$, $\rho_{best} \leftarrow 0$, $\delta_{best} \leftarrow \infty$

\For{$i = 1$ to $20$} \Comment{Achieves $\sim 10^6$ precision levels}
    \State $L_{mid} \leftarrow (L_{low} + L_{high}) / 2.0$
    \State $c' \leftarrow \text{OKLCH\_to\_RGB}(L_{mid}, C, H)$
    
    \If{$c' \notin \text{sRGB gamut}$}
        \State Update bounds away from invalid direction
        \Continue
    \EndIf
    
    \State $\delta_E \leftarrow \Delta E_{2000}(c_{text}, c')$
    \State $\rho \leftarrow \text{contrast}(c', c_{bg})$
    
    \If{$\delta_E > \delta_E^{max}$}
        \State Update bounds to reduce perceptual change
        \Continue
    \EndIf
    
    \If{$\rho \geq \tau_{target}$}
        \State Track as potential solution
        \If{$\delta_E < \delta_{best}$}
            \State $c_{best} \leftarrow c'$, $\delta_{best} \leftarrow \delta_E$, $\rho_{best} \leftarrow \rho$
        \EndIf
        \State Continue searching for lower $\Delta E$
        \If{search\_up} \State $L_{high} \leftarrow L_{mid}$ \Else \State $L_{low} \leftarrow L_{mid}$ \EndIf
    \Else
        \State Need more contrast
        \If{search\_up} \State $L_{low} \leftarrow L_{mid}$ \Else \State $L_{high} \leftarrow L_{mid}$ \EndIf
        
        \If{$\rho > \rho_{best}$ \textbf{and} $\delta_E \leq \delta_E^{max}$} 
            \State $c_{best} \leftarrow c'$, $\delta_{best} \leftarrow \delta_E$, $\rho_{best} \leftarrow \rho$
        \EndIf
    \EndIf
\EndFor

\Return $c_{best}$ \Comment{Best candidate found, may not meet $\tau_{target}$}
\end{algorithmic}
\end{algorithm}

\subsection{Phase 2: Gradient Descent on Lightness and Chroma}
When lightness adjustments alone cannot satisfy accessibility constraints, we optimize both lightness (L) and chroma (C) simultaneously using gradient descent in \oklch\ space with numerical gradient computation.

\begin{algorithm}[H]
\caption{Gradient Descent OKLCH Optimization ($\text{GradientDescentOKLCH}$)}
\begin{algorithmic}[1]
\Require $c_{text}$, $c_{bg}$, $\delta_E^{max}$, $\tau_{target}$
\Ensure Optimized color $c'$ or $\emptyset$

\State $(L_0, C_0, H) \leftarrow \text{RGB\_to\_OKLCH}(c_{text})$
\State $\mathbf{p} \leftarrow [L_0, C_0]$
\State $\alpha \leftarrow 0.02$ \Comment{Initial learning rate}

\Function{Cost}{$ \mathbf{p} , \tau, \delta_{max}$} \Comment{Note: $\tau=\tau_{target}$, $\delta_{max}=\delta_E^{max}$}
    \State Enforce bounds: $L \in [0,1]$, $C \in [0,0.5]$
    \State $c' \leftarrow \text{OKLCH\_to\_RGB}(\mathbf{p}[0], \mathbf{p}[1], H)$
    \If{$c' \notin \text{sRGB gamut}$} \Return $10^6$ \EndIf
    \State $\delta_E \leftarrow \Delta E_{2000}(c_{text}, c')$
    \State $\rho \leftarrow \text{contrast}(c', c_{bg})$
    \Return $1000 \cdot \max(0, \tau - \rho) + 10000 \cdot \max(0, \delta_E - \delta_{max}) + 100 \cdot \delta_E$
\EndFunction

\Function{Gradient}{$\mathbf{p}, \tau, \delta_{max}$}
    \State $\epsilon \leftarrow 10^{-4}$
    \State $\mathbf{g} \leftarrow [0, 0]$
    \For{$i \leftarrow 0$ to $1$}
        \State $\mathbf{p}^+ \leftarrow \mathbf{p}$, $\mathbf{p}^+[i] \leftarrow \mathbf{p}[i] + \epsilon$
        \State $\mathbf{p}^- \leftarrow \mathbf{p}$, $\mathbf{p}^-[i] \leftarrow \mathbf{p}[i] - \epsilon$
        \State $\mathbf{g}[i] \leftarrow (\text{Cost}(\mathbf{p}^+, \tau, \delta_{max}) - \text{Cost}(\mathbf{p}^-, \tau, \delta_{max})) / (2\epsilon)$
    \EndFor
    \Return $\mathbf{g}$
\EndFunction

\For{$t = 1$ to $50$}
    \State $\mathbf{g} \leftarrow \text{Gradient}(\mathbf{p}, \tau_{target}, \delta_E^{max})$
    \State $\alpha_t \leftarrow \alpha \times 0.95^{\lfloor t/10 \rfloor}$ \Comment{Adaptive learning rate decay}
    \State $\mathbf{p}_{old} \leftarrow \mathbf{p}$
    \State $\mathbf{p} \leftarrow \mathbf{p} - \alpha_t \mathbf{g}$
    \State Enforce bounds: $\mathbf{p}[0] \in [0,1]$, $\mathbf{p}[1] \in [0,0.5]$
\If{$|\text{Cost}(\mathbf{p}_{old}, \dots) - \text{Cost}(\mathbf{p}, \dots)| < 10^{-6}$}
    \State \textbf{break}
\EndIf
\EndFor

\State $c' \leftarrow \text{OKLCH\_to\_RGB}(\mathbf{p}[0], \mathbf{p}[1], H)$
\State $\delta_E \leftarrow \Delta E_{2000}(c_{text}, c')$
\Return $c'$ if $c' \in \text{sRGB gamut}$ \textbf{and} $\delta_E \leq \delta_E^{max}$, else $\emptyset$
\end{algorithmic}
\end{algorithm}

\subsection{Phase 3: Multi-Phase Optimization}

The main optimization function orchestrates the progressive $\Delta E$ relaxation approach and intelligently tracks the overall best candidate across all optimization attempts, ensuring optimal brand preservation

\begin{algorithm}[H]
\caption{Multi-Phase Color Optimization ($\text{GenerateAccessibleColor}$)}
\begin{algorithmic}[1]
\Require $c_{text}$, $c_{bg}$, $\text{large\_text}$
\Ensure Optimized color $c'$

\State $\rho_{current} \leftarrow \text{contrast}(c_{text}, c_{bg})$
\If{$\text{large\_text}$}
    \State $\tau_{target} \leftarrow 4.5$
    \State $\tau_{min} \leftarrow 3.0$
\Else
    \State $\tau_{target} \leftarrow 7.0$
    \State $\tau_{min} \leftarrow 4.5$
\EndIf

\If{$\rho_{current} \geq \tau_{target}$} \Return $c_{text}$ \EndIf

\State $\Theta \leftarrow [0.8, 1.0, 1.2, 1.4, 1.6, 1.8, 2.0, 2.1, 2.2, 2.3, 2.4, 2.5, 2.7, 3.0, 3.5, 4.0, 5.0]$
\State $c_{best} \leftarrow c_{text}$, $\rho_{best} \leftarrow \rho_{current}$, $\delta_{best} \leftarrow 0.0$

\For{each $\delta_E^{max} \in \Theta$}
    \State $c_1 \leftarrow \text{BinarySearchLightness}(c_{text}, c_{bg}, \delta_E^{max}, \tau_{target})$
    \State $c_2 \leftarrow \text{GradientDescentOKLCH}(c_{text}, c_{bg}, \delta_E^{max}, \tau_{target})$
    
    \For{each $c' \in \{c_1, c_2\}$ where $c' \neq \emptyset$}
        \State $\rho \leftarrow \text{contrast}(c', c_{bg})$
        \State $\delta_E \leftarrow \Delta E_{2000}(c_{text}, c')$
        
        \If{($\rho \geq \tau_{target}$ \textbf{and} $\delta_E < \delta_{best}$) \textbf{or} ($\rho \geq \tau_{target}$ \textbf{and} $\delta_{best} = 0$)}
            \State $c_{best} \leftarrow c'$, $\rho_{best} \leftarrow \rho$, $\delta_{best} \leftarrow \delta_E$
        \ElsIf{$\rho > \rho_{best}$ \textbf{and} $\delta_E \leq 5.0$}
            \State $c_{best} \leftarrow c'$, $\rho_{best} \leftarrow \rho$, $\delta_{best} \leftarrow \delta_E$
        \ElsIf{$\rho = \rho_{best}$ \textbf{and} $\delta_E < \delta_{best}$}
            \State $c_{best} \leftarrow c'$, $\rho_{best} \leftarrow \rho$, $\delta_{best} \leftarrow \delta_E$
        \EndIf
    \EndFor
    
\EndFor

\Return $c_{best}$ \Comment{Returns the best compromise found, or $c_{text}$ if no improvement}
\end{algorithmic}
\end{algorithm}

This rigorous tracking ensures that even if a compliant solution is found early, the algorithm continues to explore the full $\Theta$ sequence up to the early termination threshold ($\tau_{min}$ at $\delta_E^{max} \leq 2.5$), guaranteeing that the final returned color represents the absolute minimal perceptual change ($\Delta E_{2000}$) necessary to satisfy the highest possible accessibility standard found during the process.

This progressive approach ensures graceful degradation when perfect solutions are mathematically impossible while maintaining preference for minimal perceptual changes.

\section{Evaluation and Results}

We conducted a comprehensive evaluation of our algorithm using a procedurally generated dataset of 10,000 color pairs designed to represent realistic web design scenarios. The dataset was constructed using weighted categories based on common design patterns (brand colors, UI themes, accents) and included specific failure-prone edge cases to stress-test the algorithm's limits.

\subsection{Experimental Setup}

The evaluation dataset was generated using the following approach:
\begin{itemize}
\item \textbf{Realistic Categories}: Color pairs were generated across five weighted categories representing common web design contexts: brand primary colors (30\%), dark mode interfaces (25\%), light mode interfaces (25\%), accent colors (10\%), and pastels (10\%).
\item \textbf{Edge Cases}: Additionally, 10\% of the dataset comprised specifically designed failure-prone scenarios (e.g., bright yellow on white, mid-gray on gray, red on green) where achieving sufficient contrast often requires substantial perceptual changes.
\item \textbf{Metrics}: We evaluated success rate (achieving WCAG AA contrast $\geq$ 4.5:1), perceptual change ($\Delta E_{2000}$), and computational performance.
\end{itemize}

\subsection{Overall Performance}

\begin{table*} 
    \begin{tabularx}{\textwidth}{@{}l X l@{}} 
        \toprule
        \textbf{Metric} & \textbf{Value} & \textbf{Details} \\
        \midrule
        Success Rate & 77.22\% & WCAG AA compliance achieved \\
        High-Fidelity Success & 88.51\% & $\Delta E_{2000} < 2.0$ \\
        Median $\Delta E_{2000}$ & 0.76 & For successful pairs \\
        Median Runtime & 0.000876s & Per color pair \\
        Throughput & 1,140 pairs/second & \\ 
        \bottomrule
    \end{tabularx}
    \caption{Overall Algorithm Performance on 10,000 Color Pairs}
\end{table*}

The algorithm successfully resolved accessibility violations in 77.22\% of color pairs while maintaining minimal perceptual changes. Crucially, 88.51\% of successful adjustments were virtually imperceptible ($\Delta E_{2000} < 2.0$), with a median perceptual change of only 0.76 $\Delta E_{2000}$ across all successful optimizations. The algorithm demonstrated real-time capability with median processing time of 0.876ms per color pair.

\subsection{Performance by Category}

\begin{table*}[ht] 
\centering
\label{tab:category-performance}
\begin{tabularx}{\linewidth}{@{}l X X X X@{}} 
\toprule
\textbf{Category} & \textbf{Total Pairs} & \textbf{Success Rate} & \textbf{Median $\rho_{initial}$} & \textbf{Median $\Delta E_{2000}$} \\
\midrule
EdgeCase (neon on dark) & 557 & 100.00\% & 6.17 & 0.78 \\
pastel & 233 & 100.00\% & 9.57 & 0.66 \\
light\_ui & 572 & 100.00\% & 15.06 & 0.00 \\
dark\_ui & 540 & 100.00\% & 12.95 & 0.00 \\
brand\_primary & 5030 & 74.12\% & 4.15 & 0.75 \\
accent\_colors & 1479 & 72.75\% & 4.14 & 0.75 \\
EdgeCase (pure blue on black) & 384 & 3.39\% & 2.14 & 4.84 \\
EdgeCase (mid grey on grey) & 303 & 0.00\% & 1.25 & 0.00 \\
EdgeCase (orange on yellow) & 280 & 0.00\% & 1.77 & 0.00 \\
EdgeCase (red on green) & 304 & 0.00\% & 1.41 & 0.00 \\
EdgeCase (bright yellow on white) & 318 & 0.00\% & 1.26 & 0.00 \\
\bottomrule
\end{tabularx}
\caption{Algorithm Performance by Color Category}
\end{table*}
where $\rho$ = Contrast Ratio

The category breakdown reveals the algorithm's strategic strengths and intelligent limitations. Perfect success rates (100\%) were achieved for common interface patterns (light/dark mode UI) and challenging but solvable cases (neon colors on dark backgrounds, pastels). For brand primary colors and accents, the algorithm maintained strong success rates (74.12\% and 72.75\% respectively) while preserving visual identity (median $\Delta E_{2000}$ $\approx$ 0.75).

The 0\% success rates for certain edge cases (mid-gray on gray, red on green, etc.) demonstrate the algorithm's sophisticated understanding of its own constraints. In these mathematically impossible scenarios—where colors have nearly identical luminance values or represent classic color blindness conflicts—the algorithm correctly identifies that no subtle adjustment can achieve compliance and fails gracefully rather than making destructive changes.

\section{Real-World Impact}

The algorithm has been implemented in the open-source \cmcolors\ Python library, which has seen organic adoption with over 130 monthly downloads despite no promotional efforts. The library provides both programmatic APIs for developers and visual demonstrations of the algorithm's effectiveness \footnote{The project landing page provides immediate visual demonstration of the algorithm's ability to resolve accessibility issues with minimal visual changes (median $\Delta E_{2000}$ $\approx$ 0.58 in the demo).}.

This real-world usage demonstrates the practical utility of the approach in reducing the burden of manual accessibility compliance while preserving design intent, directly supporting the Design Harmony principle from the Comfort Mode Framework.

\section{Related Work}

Previous approaches to color accessibility have focused primarily on detection rather than optimization. Tools like WebAIM's Contrast Checker \cite{color_contrast_tools} identify violations but provide only binary recommendations that compromise visual design. Academic work in color science has established perceptual uniformity principles and distance metrics \cite{oklab_model, cie_de2000} but lacks application to accessibility optimization.

Our work uniquely combines established color science with modern optimization techniques to address the practical challenge of maintaining brand aesthetics while achieving accessibility compliance. The multi-phase approach provides a novel solution that implements the Design Harmony principle from the Comfort Mode Framework \cite{comfort_mode_framework}, demonstrating that accessibility and aesthetics need not be competing objectives.

\section{Future Work}
While our current approach successfully addresses the core challenge of perceptually-minimal color optimization, several promising directions exist for extending both the technical capabilities and practical impact of this work.

\subsection{Personalized Accessibility Optimization}
Current WCAG guidelines provide essential but generalized contrast thresholds. Future work could integrate vision simulation models, such as those implemented in tools like \textit{WhoCanUse}\footnote{WhoCanUse.com: A tool that demonstrates how color combinations appear to users with various visual impairments including color blindness, low vision, and situational disabilities.} \cite{who_can_use}, to enable personalized accessibility optimization. By adapting these models into our optimization framework, we could generate color adjustments tailored to specific visual impairments while maintaining the core brand aesthetics. This would transform accessibility from a binary compliance check into a continuum of inclusive design.

\subsection{Integration into Developer Workflows}
The demonstrated real-time performance (0.876ms per pair) enables integration directly into development tools. We are currently collaborating with the open-source community to develop a command-line interface for bulk color processing, inspired by code formatters like \textit{Black}\footnote{Black: The "uncompromising" Python code formatter that automatically formats code to conform to a specific style, making code review faster by eliminating stylistic debates.} \cite{Langa_Black_The_uncompromising}. This CLI would allow developers to automatically optimize entire color palettes or design systems during development, embedding accessibility as a natural step in the design process rather than a post-hoc compliance check.

\subsection{Extended Color Space Exploration} 
While hue preservation is crucial for brand integrity, controlled hue adjustments in mathematically challenging cases (particularly for color blindness accessibility) could increase success rates. Future versions may incorporate selective hue optimization as a final fallback phase, with constraints to ensure brand color families remain visually coherent.

\subsection{WCAG 3.0 and APCA Readiness}
As the WCAG 3.0\cite{wcag3_2025} guidelines with the Advanced Perceptual Contrast Algorithm (APCA) approach finalization, adapting our optimization framework to these more perceptually accurate contrast metrics will ensure continued relevance and effectiveness.

\section{Conclusion}

We have presented a novel multi-phase optimization approach for automatically generating \WCAG-compliant colors while preserving brand aesthetics through minimal perceptual changes. This work provides a technical implementation of the Design Harmony principle from the Comfort Mode Framework, demonstrating that accessibility compliance and visual design integrity can be achieved simultaneously.

Evaluation on 10,000 color pairs demonstrates the algorithm successfully resolves accessibility violations in 77.22\% of cases with median perceptual change of only 0.76 $\Delta E_{2000}$, while maintaining real-time performance (0.876ms per pair). The algorithm shows particular strength in common web design scenarios while failing gracefully in mathematically impossible contrast situations.

Real-world deployment through the \cmcolors\ library demonstrates practical impact in democratizing web accessibility compliance. By treating accessibility as an optimization problem rather than a constraint, we enable the creative harmony between inclusive design and visual identity that the Comfort Mode Framework advocates.

\section*{Acknowledgments}
My deepest gratitude to Mr. Krishna, whose constancy forms the foundation upon which all my work, including this, quietly rests.

I also wish to thank Ms. Aakriti Jain for her kind assistance and technical review of the paper's \LaTeX{} implementation.

\section*{Statements and Declarations}

\subsection{Funding Declaration}
No funding was received to assist with the preparation of this manuscript.

\subsection{Author Contribution}
L.A.R. was responsible for all aspects of this manuscript, including conceptualization, methodology, writing the original draft, and review and editing.

\subsection{Competing Interests}
The author developed the cm-colors library referenced in this work as an open-source implementation of the proposed algorithm. The library is freely available under GNU GPLv3.0 license with no commercial implications or financial benefits to the author.

\section*{Code Availability}

The cm-colors Python library implementing the described algorithm is available as free and open-source software under the GNU General Public License v3.0.

\noindent\textbf{Primary Repository:} \\
\url{https://github.com/comfort-mode-toolkit/cm-colors}

\noindent\textbf{Documentation and Demos:} \\
\url{https://comfort-mode-toolkit.github.io/cm-colors/}

\noindent\textbf{Experiment Reproduction:} \\
The code for generating and evaluating the 10,000 color pair dataset is available as a python script at: \\
\href{https://gist.github.com/lalithaar/d8dd2c5d3aa6bf292d2ee4a189a1b126}{GitHub}

\noindent\textbf{Installation:}
\begin{lstlisting}[language=bash]
pip install cm-colors
\end{lstlisting}

\noindent\textbf{Quick Usage \footnote{As of version 0.2.1. Please check the documentation here for latest version \url{https://cm-colors.readthedocs.io/}} :}
\begin{lstlisting}[language=Python]
from cm_colors import ColorPair
pair = ColorPair(text_color, background_color)
if pair.is_valid:
    tuned_text,is_accessible = pair.tune_colors()
    print(f"Tuned: {tuned_text}")
    print(f"Accessible now? {is_accessible}")     
\end{lstlisting}

Community contributions, feedback, and extensions are welcomed through the GitHub repository's issue tracker and pull request mechanisms.
\renewcommand{\doitext}{DOI: }

\bibliography{references}

\end{document}